\documentclass[11pt]{amsart}
\usepackage{epsfig}
\usepackage{verbatim}
\usepackage{graphicx,amssymb,amsmath,amsthm}
\usepackage{mathrsfs}
\usepackage{amssymb}
\usepackage{enumerate}
\newcommand{\R}{{\mathbb R}}
\newcommand{\Z}{{\mathbb Z}}

\newcommand{\C}{{\mathbb C}}

\newcommand{\wmod}[1]{\mbox{~(mod~$#1$)}}
\renewcommand{\eqref}[1]{(\ref{#1})}
\newcommand{\inner}[1]{\langle #1 \rangle}
\newcommand{\abs}[1]{\lvert#1\rvert}

\newcommand{\vA}{{\mathbf A}}

\newcommand{\m}{\mathfrak{m}}
\newcommand{\rank}{{\rm rank}}
\renewcommand{\span}{{\rm span}}

\newcommand{\vx}{{\mathbf x}}
\newcommand{\vy}{{\mathbf y}}

\newcommand{\va}{{\mathbf a}}

\renewcommand{\H}{{\mathbb F}}

\renewcommand{\top}{T}
\newtheorem{prop}{Proposition}[section]

\newtheorem{defi}{Definition}[section]

\newtheorem{theo}[prop]{Theorem}
\newtheorem{remark}[prop]{Remark}
\newtheorem{question}[prop]{Open question}

\newcommand{\argmin}[1]{\mathop{\rm argmin}\limits_{#1}}

\begin{document}
\baselineskip 18pt

\title[The minimal measurement number]{The minimal measurement number problem in  phase retrieval: a review of recent developments\footnote{Dedicated to Professor Renhong Wang on the Occasion of His Eightyth Birthday.} }

\author{Zhiqiang Xu}
\thanks{   Zhiqiang Xu was supported  by NSFC grant ( 11422113, 11331012, 91630203) and by National Basic Research Program of China (973 Program 2015CB856000).}
\address{LSEC, Inst.~Comp.~Math., Academy of
Mathematics and System Science,  Chinese Academy of Sciences, Beijing, 100091, P. R. China,\quad University of Chinese Academy of Sciences 19 A Yuquan Rd, Shijingshan District, Beijing, 100049, P.R.China}
\email{xuzq@lsec.cc.ac.cn}

\begin{abstract}
Phase retrieval is to recover the signals from phaseless measurements which is raised in many areas. A fundamental problem in phase retrieval is to determine the minimal measurement number $m$ so that one can recover $d$-dimensional signals from $m$ phaseless measurements. This problem attracts much attention of experts from different areas. In this paper, we  review the recent development on the minimal measurement number and also raise many interesting open questions.
\end{abstract}
\maketitle

\section{Introduction}
Suppose that $\vA=(\va_1,\ldots,\va_m)^T\subset \H^{m\times d}$ and $\vx_0\in \H^d$ where $\H$ is either $\R$ or $\C$.
We consider the linear equations $\vA\vx=\vA\vx_0$ where  $\vx\in \H^d$ is the unknown vector. One aim of linear algebra is to present the condition for $\vA$ under which the solution to $\vA\vx=\vA\vx_0$ is $\vx_0$. It is well known that the solution to  the linear equations $\vA\vx=\vA\vx_0$ is $\vx_0$  if and only if  ${\rm rank}(\vA)=d$.
Today, the nonlinear equation $\abs{\vA\vx}=\abs{\vA\vx_0}$ is raised in many areas   where $\abs{\vA\vx}=(\abs{\inner{\va_1,\vx}},\ldots,\abs{\inner{\va_m,\vx}})^T\in \H^m$. Naturally, one is also interested in presenting the condition for $\vA$ under which the solution to $\abs{\vA\vx}=\abs{\vA\vx_0}$ is unique.  To find the solution to the nonlinear equations is called {\em phase retrieval problem }, which is raised in many practical areas, such as  in  X-ray imaging, crystallography, electron microscopy and coherence theory. Beyond that, phase retrieval  has some fantastic connection  with many pure mathematic topics, such as the dimension of algebraic variety, the nonsingular bilinear form and the embedding problem in topology (see \cite{gphase}). Note that for any $c\in \H$ with $\abs{c}=1$ we have $\abs{\vA c\vx_0}=\abs{\vA\vx_0}$. We say the vector set $\{\va_1,\ldots,\va_m\}\subset \R^d$ or the matrix $\vA$ is {\em phase retrievable } if
\[
\{\vx\in \H^d: \abs{\vA\vx}=\abs{\vA\vx_0}\}=\tilde{\vx}_0:=\{c\vx_0:c\in \H, \abs{c}=1\}.
\]
In the context of phase retrieval, a fundamental problem is to present the minimal
measurement number $m$ so that there exists $\vA\in \R^{m\times d}$ which is
 phase retrievable.
 To state conveniently, we set
 $$
   \m_\H(d)\,:=\,\min\Bigl\{m: ~\text{ there exists $\vA=(\va_1,\ldots,\va_m)^T\in \H^{m\times d}$ which is phase retrievable in $\H^d$}\Bigr\}.
$$
 The aim of this paper is to review the recent developments about $\m_\H(d)$ and also raise many open questions. The rest
 of the paper is organized as follows. In Section 2, we introduce the results of $\m_\H(d)$ for $\H=\R$ and $\C$, respectively. We consider the case where $\vx_0$ is $s$-sparse in Section 3. Finally, the results about generalized phase retrieval are introduced in Section 4.

 \section{Phase Retrieval for general signals }
 \subsection{Real case}
The minimal measurement number problem with $\H=\R$ is investigated in \cite{BCE06}
 with presenting a condition for $\vA$ under which $\vA$ is phase retrievable. To this end, we set $\span(\vA):=\span(\{\va_1,\ldots,\va_m\})$ and $\vA_S:=(\va_j:j\in S)^T$ where $S\subset \{1,\ldots,m\}$. Then we have
 \begin{theo}\label{th:bce06}{\rm(\cite{BCE06})}   Let $\vA= (\va_1, \va_2, \dots, \va_m)^T\in \R^{m\times d}$.
The following properties  are equivalent:
\begin{itemize}
\item[{\rm ({\rm A})}] $\vA$ is  phase retrievable on $\R^d$;
\item[{\rm ({\rm B})}] For every subset $S\subset\{1,\ldots,m\}$, either $\span(\vA_S)=\R^d$ or  $\span(\vA_{S^c})=\R^d$.
\end{itemize}
\end{theo}
If $m\leq 2d-2$ then there exists $S_0\subset \{1,\ldots,m\}$ satisfying $\#S_0\leq d-1$ and $\#S_0^c\leq d-1$. Hence, $\span(\vA_{S_0})\neq \R^d$ and $\span(\vA_{S_0^c})\neq \R^d$.
According to Theorem \ref{th:bce06}, if $\vA$ is phase retrievable on $\R^d$ then we must have $m\geq 2d-1$.  We next show that $2d-1$ is the minimal measurement number which means that there exists $\vA\in \R^{(2d-1)\times d}$ satisfying (B) in Theorem \ref{th:bce06}. We set
\[
\vA_0=\left[
\begin{array}{ccccc}
1 & x_1 & x_1^2 & \cdots &  x_1^{d-1}  \\
1 & x_2 & x_2^2 & \cdots &  x_2^{d-1}  \\
\vdots &\vdots& \vdots&\vdots &\vdots \\
1 & x_{2d-1} & x_{2d-1}^2 & \cdots &  x_{2d-1}^{d-1}
\end{array}
\right]\in \R^{(2d-1)\times d}
\]
where $x_1,\ldots,x_{2d-1}\in \R$ are distinct with each other. A simple observation
is that $\vA_0$ have the property  (B) in theorem \ref{th:bce06} which implies that $\m_\R(d)=2d-1$.

\subsection{Complex case}
For the case where $\H=\C$, the minimal measurement number problem remains open.  In \cite{BCE06}, it was shown that $\vA=(\va_1,\ldots,\va_m)^T\in \C^{m\times d}$
is phase retrievable provided $m\geq 4d-2$ and $\va_1,\ldots,\va_m$ are $m$ generic vectors in $\C^d$. In \cite{bodmann}, a matrix $\vA\in \C^{(4d-4)\times d}$ is constructed and the authors also show the matrix $\vA$ is phase retrievable. The result presents an upper bound of  the minimal measurement number on $\C^d$, i.e.,  $\m_\C(d)\leq 4d-4$.  In \cite{BCMN}, one  investigated the minimal measurement number with employing the results
 from algebraic geometry. Note that $\abs{\inner{\va_j,\vx_0}}^2=Tr(\va_j\va_j^*\vx_0\vx_0^*)$
 where $\va_j\va_j^*, \vx_0\vx_0^*\in \C^{d\times d}$. Hence, one can recast
 the phase retrieval problem as a low rank matrix recovery problem:
 \[
 \text{find } {\mathbf X}\in \C^{d\times d} \quad{\text s.t.}\quad Tr(\va_j\va_j^* X)=Tr(\va_j\va_j^* \vx_0\vx_0^*), \rank(X)\leq 1, X^*=X.
 \]
Suppose that there exists $\vy_0\in \C^d$ with
 $\vy_0\notin \{c\vx_0: c\in \H, \abs{c}=1\}$  satisfying  $\abs{\inner{\va_j,\vy_0}}=\abs{\inner{\va_j,\vx_0}},
 j=1,\ldots,m$. Then we have $Tr(\va_j\va_j^*Q)=0, j=1,\ldots,m$, where
 $Q:=\vx_0\vx_0^*-\vy_0\vy_0^*$. Motivated by the observation, the following conclusion is
 obtained in \cite{BCMN}:
 \begin{prop} \label{pr:le}\cite{BCMN}
 Suppose that $\vA=(\va_1,\ldots,\va_m)^T\in \C^{m\times d}$. Then $\vA$ is not phase retrievable if and only if
 there exists a Hermitian matrix $Q\in \C^{d\times d}$ satisfying
 \[
 {\rm rank}(Q)\leq 2,\quad Tr(\va_j\va_j^*Q)=0,\quad j=1,\ldots,m.
 \]
 \end{prop}
 Based on Proposition \ref{pr:le}, in \cite{CEHV15},
 Conca, Edidin, Hering, and Vinzant apply the results
 about determinant variety to obtain the following theorem with showing $4d-4$ generic measurements are phase retrievable:
 \begin{theo}\cite{CEHV15}\label{th:alg}
Suppose that $\vA=(\va_1,\ldots,\va_m)^T\in \C^{m\times d}$.
\begin{enumerate}
\item If $m\geq 4d-4$ and $\va_1,\ldots,\va_m$ are $m$ generic vectors in $\C^d$, then $\vA$ is phase retrievable.
\item If $d=2^k+1,k\in \Z_+$ and $m<4d-4$, then $\vA$ is not phase retrievable.
\end{enumerate}
 \end{theo}
 Theorem \ref{th:alg} also shows $\m_\C(d)=4d-4$ provided
 $d$ is in the form of $2^k+1$. In \cite{CEHV15}, it was conjectured $\m_\C(d)=4d-4$ for any $d\in \Z_+$. According to Theorem \ref{th:alg}, the conjecture
  holds when $d=2,3,5,9,\ldots$. In \cite{V15}, Vinzant consider the case where $d=4$
  with constructing $11<12=4\times 4-4$ vectors $\va_1,\ldots,\va_{11}$. Employing the method from computational algebraic geometry, she verify the matrix $\vA=(\va_1,\ldots,\va_{11})$
  is phase retrievable by {\tt maple} code and hence disprove the $4d-4$  conjecture for the case $d=4$. We state her result as a proposition:
  \begin{prop}
  There exists a matrix $\vA\in \C^{11\times 4}$ which is phase retrievable. Hence $\m_\C(4)\leq 11$.
  \end{prop}

  On the other direction, one also considers the lower bound of the minimal measurement number. Usually, the the lower bound is obtained by the results from  the embedding of the complex projective space ${\mathbb P}{\mathbb C}^d$ in $\R^m$. The first lower bound $\m_\C(d)\geq 3d-2$ is presented in \cite{F3d2} and an alternative  lower bound
 $\m_\C(d)\geq 4d-3-2\alpha$  is presented in \cite{HMW13},
 where $\alpha$ denotes the number of $1$'s in the binary expansion of $d-1$. The result is improved in \cite{gphase}:
 \begin{theo}\cite{gphase}\label{th:clower}
Let $d>4$. Then $\m_\C(d)\geq 4d-2-2\alpha+\epsilon_\alpha$,
where $\alpha = \alpha(d-1)$ denotes the number of $1$'s in the binary expansion of $d-1$,
$$
\epsilon_\alpha= \left\{\begin{array}{cl} 2  & ~d \text{ odd},\, \alpha \equiv 3 \wmod 4\\
   1 &   ~d \text{ odd}, \,\alpha \equiv 2 \wmod 4\\
   0  & ~\text{otherwise.}
   \end{array}\right.
.
$$
\end{theo}

We list the minimal measurement number $\m_\C(d)$ for $d\in [2,9]\cap \Z$ in Table 1 which presents the exact value of $\m_\C(d)$ or an interval the $\m_\C(d)$ lying in.
\begin{table}[tbp]
\centering  
\begin{tabular}{ccccccccc}  
\hline
the dimension $d$  &2& 3 &4 &5 & 6 &7 &8 &9\\ \hline  
$\m_\C(d)$ &4 &8 &[10,\,11] & 16 & [19,\,20] & [23,\,24] & [26,\,28] &32\\        
 \hline
\end{tabular}
\caption{The minimal measurement number $\m_\C(d)$.}
\end{table}
The results for $d=2,3$ in Table 1 are firstly obtained in \cite{BCMN}. For the case where $d=4$, the lower bound is obtained by the result  $\m_\C(d)\geq 3d-2$ (see \cite{F3d2}) while the upper bound $\m_\C(4)\leq 11$ follows from the example in \cite{V15}. When $d\geq 5$, the upper bound is obtained by $\m_\C(d)\leq 4d-4$ while the lower bound follow from Theorem \ref{th:clower}. Note that $\alpha(d-1)=1$ and $\epsilon_\alpha=0$ provided $d$ is in the form of $2^k+1$. Theorem \ref{th:clower} implies the lower bound $\m_\C(d)\geq 4d-4$ provided $d=2^k+1$. Combining it with the upper bound $\m_\C(d)\leq 4d-4$, we recover  $\m_\C(d)=4d-4$ provided $d=2^k+1, k\geq 2$. According to Table 1, the first $d$ for which $\m_\C(d)$ is unknown is $4$. This leads us to consider the following open question:
\begin{question}
Does exist there $10$ vectors $\va_1,\ldots,\va_{10}$ so that $\vA=(\va_1,\ldots,\va_{10})^T\in \C^{10\times 4}$ is phase retrievable on $\C^4$?
\end{question}
According to the results mentioned before, we know $\m_\C(d)\leq 4d-4$. We already know $\m_\C(d)\neq 4d-4$ for some $d$. Hence, we are interested in the distance between $4d-4$ and $\m_\C(d)$. According to the lower bound presented in Theorem \ref{th:clower}, $4d-4-\m_\C(d)\leq O(\log_2(d))$. We are interested in whether the bound $O(\log_2d)$ is tight. Particularly, we would like to know whether $4d-4-\m_\C(d)$ is bound. We state the
question as follows:
\begin{question}
Is $
\limsup_{d\rightarrow \infty} (4d-\m_\C(d))
$
finite?
\end{question}
\begin{remark}

The generalized phase retrieval  is to recover $\vx\in \H^d$ from the measurement
$\{\vx^*\vA_j\vx\}_{j=1}^m$ where $\vA_j\in \H^{d\times d}$ and $\vA_j^*=\vA_j$
which includes the  {\em  phase retrieval by projection} as a special case
where each $\vA_j$ satisfying $\vA_j^2=\vA_j$ \cite{E15,phaseproj1,phaseproj2}.
Here, we assume that $\vA_j^*=\vA_j$.  The
generalized phase retrieval is investigated in \cite{gphase} with showing the connection among phase retrieval, nonsingular bilinear form and topology embedding (see \cite{gphase} for detail).
\end{remark}

\section{phase retrieval for sparse signals}
In practical applications, it is possible that some prior knowledge about $\vx_0$ is known. For example, in many applications, we know that the aim signal $\vx_0$ is sparse. We set
\[
\H_s^d:=\{\vx\in \H^d: \|\vx\|_0\leq s\},
\]
where $\H=\R$ or $\C$ and   $\|\vx\|_0$  denotes the number of the nonzero entries of $\vx$.
In this section, we assume that $\vx_0\in \H_s^d$. The $\vA\in \H^{m\times d}$ is said to be {\em $k$-sparse phase retrievable }if
\[
\{\vx\in\H^d: \abs{\vA \vx}=\abs{\vA\vx_0}\}\cap \H_s^d\,\,=\,\, \{c\vx_0:c\in \H,\abs{c}=1\}.
\]
In fact, $\vA$ is $k$-sparse phase retrievable if and only if the solution set to
\begin{equation}\label{eq:l0}
\min_\vx \|\vx\|_0 \quad s.t. \quad \abs{\vA\vx}=\abs{\vA\vx_0}
\end{equation}
is ${\tilde \vx}_0$.
In \cite{WaXu14}, Wang and Xu present the condition for $\vA$ under which $\vA$ is $s$-sparse phase retrievable.

\begin{theo}\cite{WaXu14}  \label{theo-2.1}
Suppose that $\vA=(\va_1,\ldots,\va_m)^T\in \R^{m\times d}$.
Assume that $\vA$ is $s$-sparse
phase retrievable on $\R^d$. Then $m \geq \min\,\{2s, 2d-1\}$.
Furthermore, the $\vA$ which contains $m \geq \min\,\{2s, 2d-1\}$ generically chosen
vectors in $\R^d$ is $s$-sparse
phase retrievable.
\end{theo}
For the complex case, the following result is obtained by Wang and Xu:
\begin{theo}\cite{WaXu14}  \label{theo-2.2}
Suppose that $\{\va_1,\ldots,\va_{4s-2}\}\subset \C^d$ are $m=4s-2$ generic vectors in $\C^d$. Then $\vA$ is $s$-sparse phase retrievable on $\C^d$.
\end{theo}

We consider the convex relaxation of (\ref{eq:l0}):
\begin{equation}\label{eq:l1}
\min_\vx \|\vx\|_1 \quad s.t. \quad \abs{\vA\vx}=\abs{\vA\vx_0}.
\end{equation}
Though the constraint condition in (\ref{eq:l1}) is non-convex, one still  develops many efficient algorithms to solve it \cite{algorithm1, algorithm2}. Hence, it is interesting to present the condition for $\vA$ under which the solution to $(\ref{eq:l1})$ is $\tilde{\vx}_0$ for any $\vx_0\in \H_s^d$. Motivated by the restricted isometry property in compressed sensing \cite{tao}, the strong restricted isometry property is defined in \cite{VX15}:

\begin{defi}\cite{VX15}
We say the matrix $\vA\in \R^{m\times d}$  satisfies the Strong Restricted Isometry Property (SRIP) of order $s$ and levels $\theta_-, \theta_+
\in (0,2)$
if
\[
\theta_-\|\vx\|_2^2 \,\,\leq \min_{S\subseteq \{1,\ldots,m\}, \# S \geq m/2}\|A_S\vx\|_2^2 \,\,\leq \max_{S\subseteq \{1,\ldots,m\}, \#S \geq m/2} \|A_S \vx\|_2^2 \leq \theta_+\|\vx\|_2^2
\]
holds for all $s$-sparse signals $\vx\in \R^d$.  Here $\vA_{S}:=[\va_j:j\in S]^\top$ denotes the sub-matrix of $\vA$ where only rows with
indices in $S$ are kept.
\end{defi}
The following theorem shows that the solution to (\ref{eq:l1}) is $\pm \vx_0$ provided $\vA$ satisfies SRIP:
\begin{theo}\cite{VX15}  \label{theo-3.3}
 Assume that $\vA\in \R^{m\times d}$ satisfies the Strong RIP of order $t\cdot s$ and levels $\theta_-,\theta_+$ with $t\geq \max\{\frac{1}{2\theta_--\theta_-^2}, \frac{1}{2\theta_+-\theta_+^2}\}$.
Then for any $s$-sparse signal $x_0\in \R^d$ we have
\begin{equation} \label{3.4}
   \argmin{\vx\in \R^d}\{\|\vx\|_1:~\abs{\vA \vx} = \abs{\vA \vx_0}\} = \{\pm \vx_0\},
\end{equation}
where $\abs{\vA\vx}:=[\abs{\inner{\va_j,\vx}}:j\in [m]]$ and $[m]:=\{1,\ldots,m\}$.
\end{theo}
According to Theorem \ref{theo-3.3}, it is useful to construct a matrix $\vA$ which satisfies SRIP. It was shown in the following theorem that the Gaussian random matrix $\vA\in \R^{m\times d}$ with $m=O(s\log(ed/s))$ satisfies SRIP with high probability:
\begin{theo}\cite{VX15}\label{th:SRIP}
Suppose that $t>1$ and $s\in \Z$ satisfying $tk\leq n$. Suppose  that $\vA\in \R^{m\times d}$ is a random Gaussian matrix
 whose entries $a_{jk}$ are
independent realizations of Gaussian random variables $a_{jk}\sim {\mathcal N}(0,1/m)$ and that $m\geq C\cdot ts\log (ed/ts))$.
Then
there exist constants $\theta_-, \theta_+$ with $0<\theta_-<\theta_+<2$, independent of t, such that
$A$ satisfies SRIP of order $t\cdot s$ and levels $\theta_-, \theta_+$ with probability $1-\exp(-cm/2)$, where $C, c>0$ are absolute constants.
\end{theo}
Combing Theorem \ref{theo-3.3} and Theorem \ref{th:SRIP}, we obtain (\ref{3.4}) holds
with high probability provided $\vA\in \R^{m\times d}$ with $m=O(s\log(ed/s))$ is
a Gaussian random matrix. The results in \cite{VX15} are extend to the case with the noise in \cite{GWX}. It is
interesting to extend the results in \cite{VX15} and \cite{GWX} to the complex case:
\begin{question}
Does exist there a matrix $\vA\in \C^{m\times d}$ with $m=O(s \log(ed/s))$ so that
\[
   \argmin{\vx\in \C^d}\{\|\vx\|_1:~\abs{\vA \vx} = \abs{\vA \vx_0}\}\,\, =\,\, \tilde{\vx}_0,
\]
holds for any $\vx_0\in \C_d^s$?
\end{question}
\begin{remark}
One is interested in whether it is possible to recover $\vx_0\in \R_s^d$ from $O(s\log(ed/s))$ measurements in polynomial time. According to results above, a possible way to answer this question
is to  design the polynomial time algorithm to solve (\ref{eq:l1}).
\end{remark}

\section{Conclusion}

We review some of the recent developments on the minimal measurement number problem in phase retrieval. To obtain these results, one employs some results and methods from algebraic geometry and topology. As said before,   phase retrieval can be considered as a special case of matrix recovery \cite{Xu15}. Hence, a generalized problem is to determine the minimal measurement number $m$ so that one can recover the matrix $Q\in \H^{d\times d}$ with ${\rm rank}(Q)\leq r$ from $m$ measurements. For the case $\H=\C$, the generalized problem is solved in \cite{Xu15} while it remains open for the case $\H=\R$. We believe the methods developed in phase retrieval are helpful to make some progress for the case $\H=\R$.


\end{document}